\documentclass[twocolumn,amsmath,amssymb,a4paper,prb,superscriptaddress,floatfix]{revtex4-2}
\usepackage[dvipdfmx]{graphicx}
\usepackage{natbib}
\usepackage{multirow}
\usepackage{amsmath}
\usepackage{bm}
\usepackage{mathrsfs}
\usepackage{mathtools}
\usepackage{physics}
\usepackage{url}
\usepackage{xcolor}
\usepackage[normalem]{ulem}
\usepackage{hyperref}

\begin{document}
\title{On-the-fly training of polynomial machine learning potentials in computing lattice thermal conductivity}
\author{Atsushi Togo}
\email[\phantomsection]{Author to whom any correspondence should be addressed. togo.atsushi@nims.go.jp.}
\affiliation{Center for Basic Research on Materials
  National Institute for Materials
  Science, Tsukuba, Ibaraki 305-0047, Japan}
\author{Atsuto Seko}
\affiliation{Department of Materials Science and
  Engineering, Kyoto University, Sakyo, Kyoto 606-8501, Japan}

\begin{abstract}
The application of first-principles calculations for predicting lattice thermal
conductivity (LTC) in crystalline materials, in conjunction with the linearized
phonon Boltzmann equation, has gained increasing popularity. In this
calculation, the determination of force constants through first-principles
calculations is critical for accurate LTC predictions. For material exploration,
performing first-principles LTC calculations in a high-throughput manner is now
expected, although it requires significant computational resources. To reduce
computational demands, we integrated polynomial machine learning potentials
on-the-fly during the first-principles LTC calculations. This paper presents a
systematic approach to first-principles LTC calculations. We designed and
optimized an efficient workflow that integrates multiple modular software
packages. We applied this approach to calculate LTCs for 103 compounds of the
wurtzite, zincblende, and rocksalt types to evaluate the performance of the
polynomial machine learning potentials in LTC calculations. We demonstrate a
significant reduction in the computational resources required for the LTC
predictions.
\end{abstract}
\maketitle
\section{Introduction}
\label{sec:introduction}

Calculations of lattice thermal conductivity (LTC) based on first-principles
calculations and the linearized phonon Boltzmann
equation~\cite{Peierls-Boltzmann-1929, Peierls-Quantum-Theory-of-Solids,
Ziman-electrons-phonons, Physics-of-phonons} have become increasingly popular in
recent years. This is because sufficiently accurate LTC values can be
systematically predicted for a wide variety of crystals using available computer
simulation packages.\cite{ALAMODE, almaBTE, PhonTS,
Hellman-TDEP-2013, phono3py, ShengBTE_2014} These
computational tools are expected to be applied in materials discovery within a
high-throughput calculation environment. However, since first-principles LTC
calculations are still computationally intensive, there is a need for the
development of methodologies to reduce the computational demands.

We conventionally employ a supercell approach combined with the finite displacement
method for first-principles LTC calculations. Random or systematic displacements
are introduced to the supercells, and the forces on atoms are calculated using
first-principles calculations. Subsequently, supercell force constants are
computed from the dataset composed of the displacements and forces, and the LTC
values are calculated from these supercell force constants. Many supercells with
different displacement configurations are often required to populate the tensor
elements of the supercell force constants.

The accuracy of predicting LTCs relies on the use of first-principles
calculations to obtain the displacement-force dataset. However, this approach is
computationally intensive. In order to achieve precise LTC predictions with
lower computational resources, compressive sensing force constants calculation
methods were developed, as reported in
Refs.~\onlinecite{Zhou-PRL-compressive-sensing-FC-2014, Tadano-2015}. These
methods employ regularized linear regression techniques to eliminate certain
tensor elements of the supercell force constants, thereby reducing the required
size of the displacement-force dataset.

In this study, we introduce another approach to reduce the computational demands
of first-principles LTC calculations. We incorporate polynomial machine learning
potentials (MLPs)~\cite{polymlp, pypolymlp} into an intermediate stage of the
LTC calculation process. The polynomial MLPs are trained using a small dataset
of displacement-force pairs and energies derived from first-principles
calculations. Subsequently, the polynomial MLPs generate a large
displacement-force dataset to calculate supercell force constants with
significantly lower computational demands than those required by
first-principles calculations. Compared to the compressive sensing approach,
this approach may more efficiently represent energy surface of crystal
potential. However, the workflow for performing the first-principles LTC
calculation with this approach is more complex. One objective of this study is
to encapsulate this intricate workflow into a software package, complete with a
set of well-optimized default parameters. Developing this software package
requires a thorough understanding of the MLP code details. Moreover, efficient
estimations of polynomial MLPs can be accomplished using linear regressions
supported by powerful libraries for linear algebra. Therefore, we have utilized
the polynomial MLP code developed by one of the authors.

The computational procedure is illustrated in Fig.~\ref{fig:software-chain}.
Initially, a set of supercells with random displacements of atoms is prepared.
Forces on atoms and energies in the supercells are calculated using
first-principles calculations. The dataset, consisting of displacement-force
pairs and energies, is employed to train the polynomial MLPs. Forces on atoms in
another set of supercells with random displacements of atoms are calculated
using the trained polynomial MLPs. Supercell force constants are then calculated
from the displacement-force dataset obtained through the polynomial MLPs.
Finally, the LTC values are calculated using the supercell force constants
obtained. Crystal symmetry plays an important role in reducing the computational
demands and improving the numerical accuracy.

The goal of the methodological and software developments presented in this study
is to reduce the computational demands of first-principles LTC calculations,
with the aim of high-throughput LTC calculations. At the same time, we
prioritize user convenience, considering factors such as calculation time and
required memory. We have designed the workflow and software to make the use of
the polynomial MLPs appear straightforward from the user's perspective. This
study explores the feasibility of employing polynomial MLPs as an intermediate
stage in the calculation of third-order supercell force constants for
first-principles LTC calculations.

In this study, a systematic calculation of LTCs at 300 K was performed for the
same set of the 103 compounds of wurtzite, zincblende, and rocksalt types
reported in Ref.~\onlinecite{phonopy-phono3py-JPSJ}. The computational
workflow-design and details are presented in Secs.~\ref{sec:design-of-workflow} and
\ref{sec:computational-method}. While most of the theoretical and methodological
background is covered in the referenced articles, those relevant to this paper
are described in Sec.~\ref{sec:computational-method}. The results of the LTC
calculations are summarized in Sec.~\ref{sec:results-and-discussion}.

\section{Design of computational workflow}
\label{sec:design-of-workflow}

\begin{figure}[ht]
  \begin{center}
    \includegraphics[width=1.00\linewidth]{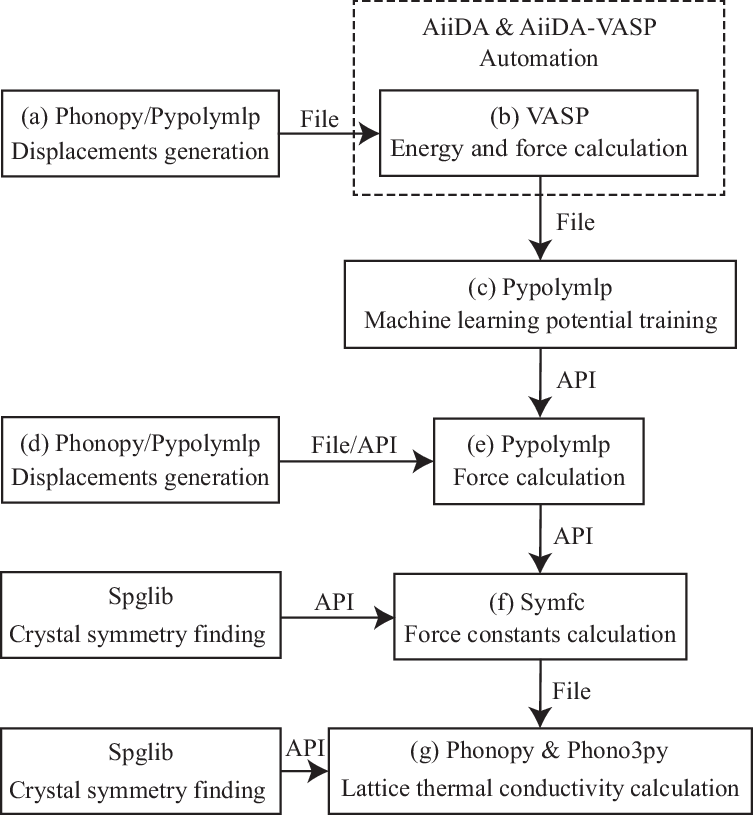}
    \caption{\label{fig:software-chain}
      Schematic illustration of workflow employed in this study for LTC
      calculations. Each calculation step is represented by a box, within which
      the name of the software package and the type of calculation are
      described. The arrows roughly indicate flow of data. The data were passed
      via APIs or computer files. The steps are as follows: Step (a): Generation
      of a modest number of supercells with random displacements. Step (b):
      Calculation of energies and forces in the supercells generated in step (a)
      by first-principles calculations. The submission of a large number of
      computational jobs is automated using a workflow system. Step (c):
      Training of polynomial MLPs using the energies and forces in the
      supercells obtained in step (b). Step (d): Generation of a large number of
      supercells with random displacements. Step (e): Calculation of forces in
      the supercells generated in step (d) using the trained polynomial MLPs.
      Step (f): Calculation of force constants from the displacement-force
      dataset calculated in step (e). Step (g): Calculation of LTC using the
      supercell force constants obtained in step (f). This study introduces
      steps (c), (d), and (e) to the workflow. Conventionally, the dataset from
      step (b) is used directly in step (f). This conventional approach is
      computationally demanding due to the necessity of a large dataset in step
      (b). In this workflow, second-order force constants can be calculated in
      the conventional approach because the computational demand is much less
      than the calculation of third-order force constants in the conventional
      approach.}

  \end{center}
\end{figure}

As shown in Fig.~\ref{fig:software-chain}, our study utilized a specific
approach for LTC calculations. This approach consists of specialized modules for
each calculation step. These modules are interconnected through a local
file system and data communication, including computer-network file transfers
and application program interfaces (APIs).

In the LTC calculation process, we generate two distinct sets of supercells with
random displacements of atoms. This occurs in steps (a) and (d), as illustrated
in Fig.~\ref{fig:software-chain}. In this study, we displaced atoms by a
constant distance in random directions. In step (b), we conduct energy and force
calculations for the first set of supercells using first-principles
calculations. This is the most computationally demanding step. The outputs of
step (b) form a dataset used to train polynomial MLPs in step (c). In step (e),
we calculate forces for the second set of supercells using the trained
polynomial MLPs. In step (f), third-order supercell force constants are computed
using the displacement-force dataset from step (e). Finally in step (g), LTC is
calculated from the supercell force constants. Use of crystal symmetry is
important in steps (f) and (g) for the computational efficiency and numerical
accuracy. The computational demand from step (c) to the end is negligible
compared to step (b).

Our computational workflow is specifically designed to optimize high-throughput
LTC calculations, balancing efficiency and convenience. This workflow is divided
into two main parts: dataset preparation and calculations using this dataset.
The dataset preparation stage involves a set of energy and force calculations
using first-principles calculations, which are normally distributed over
computer nodes to conduct the calculations in parallel. After completing these
calculations, we extract the necessary data from the output files of the
first-principles calculations. This data is then saved for transfer via
computer-network communication. In the subsequent steps, calculations are
performed on a single computer. Depending on the operational requirements, data
transfer between the modules is facilitated either through APIs or via the local
computer file system for ease of use.

\section{Computational methods}
\label{sec:computational-method}

\subsection{LTC calculation}

LTC values were computed by solving the Peierls-Boltzmann equation within the
relaxation time approximation (RTA)\cite{Peierls-Boltzmann-1929,
Peierls-Quantum-Theory-of-Solids, Allen-LTC-2018} using the phono3py
code.\cite{phono3py, phonopy-phono3py-JPCM} It is important to note that the
method employed for the LTC calculations may not be appropriate for compounds
with low or high thermal conductivity among the 103 compounds selected. To
simplify the methodological investigation in this study, we considered only
phonon-phonon scattering for determining the phonon relaxation times. Imaginary
parts of phonon self-energies corresponding to the bubble diagram were
calculated from supercell force constants, where a linear tetrahedron method was
used to treat the delta function,\cite{phonopy-phono3py-JPCM} and their
reciprocals were used as the relaxation times. Computational details on the
calculation of the supercell force constants are provided in the subsequent
sections. Phonon group velocities, mode heat capacities, frequencies, and
eigenvectors were obtained from dynamical matrices constructed using
second-order supercell force constants. Additionally, a non-analytical term
correction~\cite{Pick-1970, Gonze-1994, Gonze-1997} was applied to the dynamical
matrices to account for long-range dipole-dipole interactions in the harmonic
phonon calculation. For the wurtzite-type compounds, reciprocal spaces were
sampled using a $19 \times 19 \times 10$ mesh, while a $19 \times 19 \times 19$
mesh was employed for the zincblende-type and rocksalt-type compounds.

\subsection{Supercell force constants calculation}
Force constants required for predicting LTCs are determined as the coefficients
$\Phi_{l\kappa\alpha,\ldots}$ in the Taylor expansion of the potential energy
$\mathscr{V}$ with respect to atomic displacements $u_{l\kappa\alpha}$:
\begin{align}
  \label{eq:potential-expansion}
  \mathscr{V} =
        \Phi_0 & + \sum_{l\kappa\alpha} \Phi_{l\kappa\alpha}
  u_{l\kappa\alpha} \nonumber                                                             \\
  +             & \frac{1}{2}\sum_{l\kappa\alpha, l'\kappa'\alpha'}
  \Phi_{l\kappa\alpha,l'\kappa'\alpha'}
  u_{l\kappa\alpha} u_{l'\kappa'\alpha'} \nonumber                                        \\
  +             & \frac{1}{3!}\sum_{l\kappa\alpha, l'\kappa'\alpha', l''\kappa''\alpha''}
  \nonumber                                                                               \\
                &
  \Phi_{l\kappa\alpha, l'\kappa'\alpha', l''\kappa''\alpha''}
  u_{l\kappa\alpha} u_{l'\kappa'\alpha'} u_{l''\kappa''\alpha''}
  + \cdots,
\end{align}
where $l$, $\kappa$, and $\alpha$ represent the unit cell, atom index within the
unit cell, and Cartesian coordinate, respectively. By differentiating both sides
of Eq.~(\ref{eq:potential-expansion}) with respect to $u_{l\kappa\alpha}$, we
obtain
\begin{align}
  \label{eq:fitting-equation}
  -f_{l\kappa\alpha} = &
  \Phi_{l\kappa\alpha} +
  \sum_{l'\kappa'\alpha'}
  \Phi_{l\kappa\alpha,l'\kappa'\alpha'}
  u_{l'\kappa'\alpha'} \nonumber                                        \\
  +             & \frac{1}{2}\sum_{l'\kappa'\alpha', l''\kappa''\alpha''}
  \Phi_{l\kappa\alpha, l'\kappa'\alpha', l''\kappa''\alpha''}
  u_{l'\kappa'\alpha'} u_{l''\kappa''\alpha''}  \nonumber \\
  + & \cdots,
\end{align}
where $f_{l\kappa\alpha}$ represents the $\alpha$-component of the force on atom
$l\kappa$. In this study, we obtain these coefficients by fitting a dataset
consisting of finite displacements and forces of atoms in supercells
approximating Eq.~(\ref{eq:fitting-equation}). The equation that we employ for
fitting the third-order supercell force constants is written as
\begin{align}
  \label{eq:fitting-equation-sc}
  -F_{l\kappa\alpha} = &
  \sum_{l'\kappa'\alpha'}
  \Phi^\text{SC}_{l\kappa\alpha,l'\kappa'\alpha'}
  U_{l'\kappa'\alpha'} \nonumber                                        \\
  +             & \frac{1}{2}\sum_{l'\kappa'\alpha', l''\kappa''\alpha''}
  \Phi^\text{SC}_{l\kappa\alpha, l'\kappa'\alpha', l''\kappa''\alpha''}
  U_{l'\kappa'\alpha'} U_{l''\kappa''\alpha''},
\end{align}
where $F_{l\kappa\alpha}$ and $U_{l\kappa\alpha}$ represent the forces and
displacements of atoms in supercells, respectively.
$\Phi^\text{SC}_{l\kappa\alpha,l'\kappa'\alpha'}$ and
$\Phi^\text{SC}_{l\kappa\alpha,l'\kappa'\alpha',l''\kappa''\alpha''}$ denote the
second- and third-order supercell force constants, respectively. Here we assume
$\Phi^\text{SC}_{l\kappa\alpha}=0$.

Higher-order terms are effectively included in
$\Phi^\text{SC}_{l\kappa\alpha,l'\kappa'\alpha'}$ and
$\Phi^\text{SC}_{l\kappa\alpha,l'\kappa'\alpha',l''\kappa''\alpha''}$. In
Eq.~(\ref{eq:fitting-equation-sc}), the contributions from higher-order terms
are expected to become negligible as the displacements become smaller. However,
use of smaller displacements can make the computation of supercell force
constants more susceptible to numerical errors of forces. To find a compromise
between these conflicting requirements, a modest inclusion of higher-order
contributions is commonly adopted.  Higher-order terms also introduce additional
degrees of freedom. To average over them in the third-order supercell force
constants, a larger displacement-force dataset is required to achieve
convergence in LTC values.

The linear regression method was employed to calculate supercell force constants
from the displacement-force dataset. In this method, forces acting on atoms,
which were randomly displaced from their equilibrium positions in supercells,
were computed either using the polynomial MLPs in our current approach or
through first-principles calculations in the conventional approach. Tensor
elements of supercell force constants were projected onto the subspace defined
by symmetry projection operators of totally symmetric irreducible
representations of the space group, index permutation, and translational
invariance. In addition, detailed techniques were developed to enhance the
efficiency of this computational process. This process is implemented in the
symfc code.\cite{symfc-project,symfc} The crystallographic symmetries were
determined using the spglib code.\cite{spglib}

For the zincblende- and rocksalt-type compounds, we utilized supercells with
$2\times 2\times 2$ and $4\times 4\times 4$ expansions of the conventional unit
cells to calculate third-order and second-order supercell force constants,
respectively. In the case of the wurtzite-type compounds, supercells with
$3\times 3\times 2$ and $5\times 5\times 3$ expansions of the unit cells were
employed for third-order and second-order supercell force constants
calculations, respectively.

To compute the second-order supercell force constants, we employed the finite
difference method as implemented in the phonopy
code.\cite{phonopy-phono3py-JPCM,phonopy-phono3py-JPSJ} We used the same
displacement-force datasets as those in Ref.~\onlinecite{phonopy-phono3py-JPSJ},
where the forces in these datasets had been computed through first-principles
calculations. The number of supercells in the datasets were six, two, and two
for the wurtzite-, zincblende-, and rocksalt-type compounds, respectively.

Third-order supercell force constants were calculated from the
displacement-force datasets using the symfc code.\cite{symfc-project} The
numbers of symmetrically independent force constant elements were 7752, 1536,
and 758 for the wurtzite-, zincblende-, and rocksalt-type compounds,
respectively. These values were determined based on the forces acting on atoms,
which were inferred using the polynomial MLPs implemented in the pypolymlp
code.\cite{pypolymlp}

\subsection{Polynomial MLPs}
The polynomial MLPs were trained using the dataset composed of forces and
displacements of atoms and energies in supercells. These energies and forces
were computed through first-principles calculations. The performance of the
polynomial MLPs in the LTC calculation via the third-order supercell force
constants is discussed in Section~\ref{sec:results-and-discussion}.

For the 103 compounds, we trained the polynomial MLPs using the pypolymlp
code.\cite{pypolymlp} In this training, Gaussian-type radial functions were
employed, and the functional form $f_n(r)$ is given as
\begin{align}
  \label{eq:gaussian-type-rad-func}
  f_n(r)        & = \exp[-\beta_n(r-r_n)^2] f_\text{c}(r), \\
  f_\text{c}(r) & =
  \begin{cases}
    [\cos(\pi r/r_\text{c}) + 1] / 2 \;\; (r \leq r_\text{c}), \\
    0 \;\; (r > r_\text{c}).
  \end{cases}
\end{align}
where $r$ represents the distance from the center of each atom, and $r_\text{c}$
is the cutoff distance. $\beta_n$ and $r_n$ are the parameters, respectively.
Radial functions with $r_\text{c}=8.0$ \AA~and $\beta_n=1.0$ \AA$^{-2}$ and
$r_n=(n-1) (r_\text{c}-1.0)/11$ for $n=1,\ldots,12$ were used. We considered
polynomial invariants up to third order characterizing neighboring atomic
density based on spherical harmonics with the maximum angular numbers of
spherical harmonics $l_\text{max}^{(2)}=l_\text{max}^{(3)}=8$. The polynomial
models were then constructed by the polynomial functions of the pair invariants
and linear polynomial function of the polynomial invariants. We considered
polynomial functions up to second order. The model coefficients were estimated
from electronic total energies and forces by the linear ridge regression method.

\subsection{First-principles calculation}

For the first-principles calculations, we employed the plane-wave basis
projector augmented wave (PAW) method~\cite{PAW-Blochl-1994} within the
framework of DFT as implemented in the VASP
code.\cite{VASP-Kresse-1995,VASP-Kresse-1996,VASP-Kresse-1999} The generalized
gradient approximation (GGA) of Perdew, Burke, and Ernzerhof revised for solids
(PBEsol)~\cite{PBEsol} was used as the exchange correlation potential. To ensure
high numerical accuracy in computing atomic forces, the projection operators
were applied in reciprocal spaces and additional support grids were employed for
the evaluation of the augmentation charges. Static dielectric constants and Born
effective charges were calculated with the conventional unit cells from density
functional perturbation theory (DFPT) as implemented in the VASP
code.\cite{Gajdos-2006,Wu-2005}

A plane-wave energy cutoff of 520 eV was employed for the supercell force
calculations and 676 eV for the DFPT calculations. Reciprocal spaces of the
zincblende- and rocksalt-type compounds were sampled by the half-shifted
$2\times 2\times 2$ meshes for the $2\times 2\times 2$ supercells, the
half-shifted $1\times 1\times 1$ meshes for the $4\times 4\times 4$ supercells,
and the half-shifted $8\times 8 \times 8$ meshes for the conventional unit
cells. Reciprocal spaces of the wurtzite-type compounds were sampled by the
$2\times 2\times 2$ meshes that are half-shifted along the $c^*$ axis for the
$3\times 3\times 2$ supercells, the $1\times 1\times 2$ meshes that are
half-shifted along the $c^*$ axis for the $5\times 5\times 3$ supercells, and
the $12\times 12 \times 8$ meshes that are half-shifted along the $c^*$ axis for
the unit cells.

\subsection{Automation of dataset preparation}

Performing a large number of first-principles calculations can be
computationally intensive and may require high-performance computing resources.
This stage consumes a significant amount of computational power throughout the
LTC calculation process. It is virtually inevitable that some of these
calculations fail for various reasons, such as reaching the maximum number of
electronic structure convergence iterations or encountering issues related to
computer networks and hardware. Although the proportion of failed calculations
was relatively low, we have not yet fully automated error recovery for all
possible cases.

We systematically identified calculation failures and re-executed those
calculations semi-manually with the assistance of the workflow system instead of
attempting to fully automate all processes. After completing all the supercell
calculations using first-principles calculations, the dataset for each compound
required for the subsequent LTC calculation process was composed into a single
computer file in a structured format.

For the systematic calculations of energies and forces in supercells using
first-principles calculations, we utilized the AiiDA
environment~\cite{AiiDA,AiiDA-Huber-2020,AiiDA-Uhrin-2021} in conjunction with
the AiiDA-VASP plugin.\cite{AiiDA-VASP} The advantage of using the workflow
automation system was not only the automation of submitting calculation jobs to
high-performance computers, but also the automated data storing of the
calculation results in a database systematically. The computed data, stored
within the AiiDA database, could be conveniently accessed through the Python
programming language. By writing a concise Python script, we were able to
extract supercell energies, forces, and displacements from the AiiDA database on
demand and convert this data into the structured format required for immediate
use by the phono3py code.\cite{phono3py, phonopy-phono3py-JPCM} An example of
the phono3py data format can be found in the phono3py github repository.

\subsection{Parameters for 103 binary compounds}
33 compounds for the wurtzite- and zincblende-type and 37 compounds for the
rocksalt-type were used to evaluate the LCT calculation approach proposed in
this study, and their chemical compositions are listed in Tables
\ref{table:wurtzite} and \ref{table:rocksalt}. Crystal structures of the
wurtzite and zincblende types are similar, though their stacking orders are
different, much like the relationship between face-centered-cubic and
hexagonal-close-packed structure types. Since it is of interest to explore their
similarities and differences in calculations, as also studied in
Ref.~\onlinecite{phono3py}, the compounds with the same chemical compositions
for the wurtzite and zincblende types were calculated. The tables also provide
information on lattice parameters, the choices of PAW datasets from the VASP
package, and electronic total energies of the elements that were subtracted from
the total energies of the compounds used to train the polynomial MLPs.

\begin{table}[ht]
  \caption{\label{table:wurtzite} Lattice parameters, names of the VASP PAW-PBE
  datasets, electronic total energies of the atoms used in this study for 33
  wurtzite- and zincblende-type compounds. w-$a$, w-$c$, and z-$a$ denote the
  lattice parameters $a$ and $c$ of the wurtzite-type compounds, and $a$ of the
  zincblende-type compounds, respectively.}
  \begin{ruledtabular}
    \begin{tabular}{lcccrcrc}
      & w-$a$ & w-$c$ & z-$a$ & \multicolumn{2}{c}{energy (eV)} & \multicolumn{2}{c}{energy (eV)} \\
      \hline
      AgI  & 4.56 & 7.45 & 6.44 & (Ag\_pv) & $-0.233$ & (I)      & $-0.182$ \\
      AlAs & 4.00 & 6.58 & 5.67 & (Al)     & $-0.282$ & (As\_d)  & $-0.989$ \\
      AlN  & 3.11 & 4.98 & 4.38 & (Al)     & $-0.282$ & (N)      & $-1.905$ \\
      AlP  & 3.86 & 6.34 & 5.47 & (Al)     & $-0.282$ & (P)      & $-1.140$ \\
      AlSb & 4.35 & 7.16 & 6.17 & (Al)     & $-0.282$ & (Sb)     & $-0.828$ \\
      BAs  & 3.35 & 5.55 & 4.77 & (B)      & $-0.359$ & (As\_d)  & $-0.989$ \\
      BeO  & 2.70 & 4.38 & 3.80 & (Be)     & $-0.023$ & (O)      & $-0.957$ \\
      BeS  & 3.41 & 5.63 & 4.84 & (Be)     & $-0.023$ & (S)      & $-0.578$ \\
      BeSe & 3.62 & 5.97 & 5.14 & (Be)     & $-0.023$ & (Se)     & $-0.438$ \\
      BeTe & 3.95 & 6.53 & 5.61 & (Be)     & $-0.023$ & (Te)     & $-0.359$ \\
      BN   & 2.54 & 4.20 & 3.61 & (B)      & $-0.359$ & (N)      & $-1.905$ \\
      BP   & 3.18 & 5.27 & 4.52 & (B)      & $-0.359$ & (P)      & $-1.140$ \\
      CdS  & 4.13 & 6.72 & 5.84 & (Cd)     & $-0.021$ & (S)      & $-0.578$ \\
      CdSe & 4.31 & 7.03 & 6.09 & (Cd)     & $-0.021$ & (Se)     & $-0.438$ \\
      CdTe & 4.59 & 7.52 & 6.50 & (Cd)     & $-0.021$ & (Te)     & $-0.359$ \\
      CuBr & 3.92 & 6.48 & 5.56 & (Cu\_pv) & $-0.274$ & (Br)     & $-0.225$ \\
      CuCl & 3.70 & 6.17 & 5.27 & (Cu\_pv) & $-0.274$ & (Cl)     & $-0.311$ \\
      CuH  & 2.81 & 4.44 & 3.93 & (Cu\_pv) & $-0.274$ & (H)      & $-0.946$ \\
      CuI  & 4.17 & 6.88 & 5.92 & (Cu\_pv) & $-0.274$ & (I)      & $-0.182$ \\
      GaAs & 3.99 & 6.57 & 5.66 & (Ga\_d)  & $-0.286$ & (As\_d)  & $-0.989$ \\
      GaN  & 3.18 & 5.18 & 4.50 & (Ga\_d)  & $-0.286$ & (N)      & $-1.905$ \\
      GaP  & 3.83 & 6.31 & 5.44 & (Ga\_d)  & $-0.286$ & (P)      & $-1.140$ \\
      GaSb & 4.31 & 7.10 & 6.11 & (Ga\_d)  & $-0.286$ & (Sb)     & $-0.828$ \\
      InAs & 4.30 & 7.05 & 6.09 & (In\_d)  & $-0.264$ & (As\_d)  & $-0.989$ \\
      InN  & 3.54 & 5.71 & 4.99 & (In\_d)  & $-0.264$ & (N)      & $-1.905$ \\
      InP  & 4.15 & 6.81 & 5.88 & (In\_d)  & $-0.264$ & (P)      & $-1.140$ \\
      InSb & 4.60 & 7.56 & 6.52 & (In\_d)  & $-0.264$ & (Sb)     & $-0.828$ \\
      MgTe & 4.56 & 7.41 & 6.44 & (Mg\_pv) & $-0.009$ & (Te)     & $-0.359$ \\
      SiC  & 3.08 & 5.05 & 4.36 & (Si)     & $-0.522$ & (C)      & $-1.340$ \\
      ZnO  & 3.24 & 5.23 & 4.56 & (Zn)     & $-0.016$ & (O)      & $-0.957$ \\
      ZnS  & 3.79 & 6.21 & 5.36 & (Zn)     & $-0.016$ & (S)      & $-0.578$ \\
      ZnSe & 3.98 & 6.54 & 5.64 & (Zn)     & $-0.016$ & (Se)     & $-0.438$ \\
      ZnTe & 4.28 & 7.05 & 6.07 & (Zn)     & $-0.016$ & (Te)     & $-0.359$ \\
    \end{tabular}
  \end{ruledtabular}
\end{table}

\begin{table}[ht]
  \caption{\label{table:rocksalt} Lattice parameters $a$, names of the VASP
  PAW-PBE datasets, and electronic total energies of the atoms used in this
  study for 37 rocksalt-type compounds.}
  \begin{ruledtabular}
    \begin{tabular}{lcrcrc}
      & $a$ & \multicolumn{2}{c}{energy (eV)} & \multicolumn{2}{c}{energy (eV)} \\
      \hline
      AgBr & 5.67 & (Ag\_pv) & $-0.233$ & (Br)     & $-0.225$ \\
      AgCl & 5.44 & (Ag\_pv) & $-0.233$ & (Cl)     & $-0.311$ \\
      BaO  & 5.53 & (Ba\_sv) & $-0.035$ & (O)      & $-0.957$ \\
      BaS  & 6.36 & (Ba\_sv) & $-0.035$ & (S)      & $-0.578$ \\
      BaSe & 6.58 & (Ba\_sv) & $-0.035$ & (Se)     & $-0.438$ \\
      BaTe & 6.97 & (Ba\_sv) & $-0.035$ & (Te)     & $-0.359$ \\
      CaO  & 4.77 & (Ca\_pv) & $-0.010$ & (O)      & $-0.957$ \\
      CaS  & 5.63 & (Ca\_pv) & $-0.010$ & (S)      & $-0.578$ \\
      CaSe & 5.87 & (Ca\_pv) & $-0.010$ & (Se)     & $-0.438$ \\
      CaTe & 6.30 & (Ca\_pv) & $-0.010$ & (Te)     & $-0.359$ \\
      CdO  & 4.71 & (Cd)     & $-0.021$ & (O)      & $-0.957$ \\
      CsF  & 5.96 & (Cs\_sv) & $-0.166$ & (F)      & $-0.556$ \\
      KBr  & 6.59 & (K\_pv)  & $-0.182$ & (Br)     & $-0.225$ \\
      KCl  & 6.29 & (K\_pv)  & $-0.182$ & (Cl)     & $-0.311$ \\
      KF   & 5.37 & (K\_pv)  & $-0.182$ & (F)      & $-0.556$ \\
      KH   & 5.63 & (K\_pv)  & $-0.182$ & (H)      & $-0.946$ \\
      KI   & 7.05 & (K\_pv)  & $-0.182$ & (I)      & $-0.182$ \\
      LiBr & 5.41 & (Li\_sv) & $-0.286$ & (Br)     & $-0.225$ \\
      LiCl & 5.06 & (Li\_sv) & $-0.286$ & (Cl)     & $-0.311$ \\
      LiF  & 4.00 & (Li\_sv) & $-0.286$ & (F)      & $-0.556$ \\
      LiH  & 3.97 & (Li\_sv) & $-0.286$ & (H)      & $-0.946$ \\
      LiI  & 5.90 & (Li\_sv) & $-0.286$ & (I)      & $-0.182$ \\
      MgO  & 4.22 & (Mg\_pv) & $-0.009$ & (O)      & $-0.957$ \\
      NaBr & 5.93 & (Na\_pv) & $-0.246$ & (Br)     & $-0.225$ \\
      NaCl & 5.60 & (Na\_pv) & $-0.246$ & (Cl)     & $-0.311$ \\
      NaF  & 4.63 & (Na\_pv) & $-0.246$ & (F)      & $-0.556$ \\
      NaH  & 4.79 & (Na\_pv) & $-0.246$ & (H)      & $-0.946$ \\
      NaI  & 6.41 & (Na\_pv) & $-0.246$ & (I)      & $-0.182$ \\
      PbS  & 5.90 & (Pb\_d)  & $-0.374$ & (S)      & $-0.578$ \\
      PbSe & 6.10 & (Pb\_d)  & $-0.374$ & (Se)     & $-0.438$ \\
      PbTe & 6.44 & (Pb\_d)  & $-0.374$ & (Te)     & $-0.359$ \\
      RbBr & 6.88 & (Rb\_pv) & $-0.168$ & (Br)     & $-0.225$ \\
      RbCl & 6.58 & (Rb\_pv) & $-0.168$ & (Cl)     & $-0.311$ \\
      RbF  & 5.66 & (Rb\_pv) & $-0.168$ & (F)      & $-0.556$ \\
      RbH  & 5.95 & (Rb\_pv) & $-0.168$ & (H)      & $-0.946$ \\
      RbI  & 7.32 & (Rb\_pv) & $-0.168$ & (I)      & $-0.182$ \\
      SrO  & 5.13 & (Sr\_sv) & $-0.032$ & (O)      & $-0.957$ \\
    \end{tabular}
  \end{ruledtabular}
\end{table}

\section{Results}
\label{sec:results-and-discussion}

\subsection{Choice of displacements and number of supercells}
\label{sec:choice-of-disps}

For each compound, two distinct displacement-force datasets that share the same
supercell basis vectors were employed to calculate LTCs. Energies and forces of
the supercells in the first dataset were computed using first-principles
calculations, while the polynomial MLPs were utilized for calculating forces in
the second dataset. The first dataset was used to train the polynomial MLPs. The
second dataset was employed to compute third-order supercell force constants
by fitting.

To investigate the performance of the polynomial MLPs in predicting LTC values,
100 supercells with random directional displacements were initially prepared as
the first dataset. Subsequently, the first 10, 20, 40, 60, and 80 supercells
were selected from the list of 100 supercells as subsets. Using the
displacement-force pairs and energies of these supercells, the polynomial MLPs
were trained, and the last 20 supercells were reserved as test data to optimize
their ridge regularization parameters.

For the ease of use of the software package, we decided to employ a constant
displacement distance, and to obtain reasonable LTC values, we chose a constant
displacement distance of 0.03 \AA. Interestingly, we found that the polynomial
MLPs performed well even with a relatively large displacement distance, such as
0.1 \AA. It is important to note that these factors are highly dependent on the
specific force calculators and calculation configurations used.

We utilized another displacement-force dataset that consists of 400 supercells
with random directional displacements for the computation of third-order
supercell force constants. These supercell forces were calculated using the
trained polynomial MLPs, where the residual forces were subtracted. The
root-mean-square errors of the polynomial MLPs trained on the 20 supercells
ranged from approximately
% 0.00000549136772071204 to 0.0014214735438958888
$5.5 \times 10^{-6}$ to $1.4 \times 10^{-3}$ eV/\AA, which are expected to
represent the same degree of numerical errors in the displacement-force dataset.

Due to the numerical smoothness of the polynomial MLPs for the force calculation
with respect to positions of atoms compared to the first-principles calculations
employed in this study, we were able to choose a small constant displacement
distance of 0.001 \AA. This benefits better convergence with smaller dataset
when fitting the supercell force constants by
Eq.~(\ref{eq:fitting-equation-sc}). For instance, in the case of a displacement
distance of 0.03 \AA, it was necessary to employ 10000 supercells to achieve
well converged LTC values for the 103 compounds. This suggests that when
high-order force constants are more relevant for specific compounds, direct
calculation of third-order supercell force constants from the the
displacement-force dataset through first-principles calculations may require a
large dataset to achieve convergence of LTC values.

% For strongly anharmonic crystals, self-consistent phonon methods, which were not
% considered in this study, may have to be employed to obtain physically more
% meaningful results. In such cases, the use of the polynomial MLPs can also be
% beneficial to accelerate the LTC calculations.

\subsection{Calculated LTCs}

\begin{figure*}[ht]
  \begin{center}
    \includegraphics[width=1.0\linewidth]{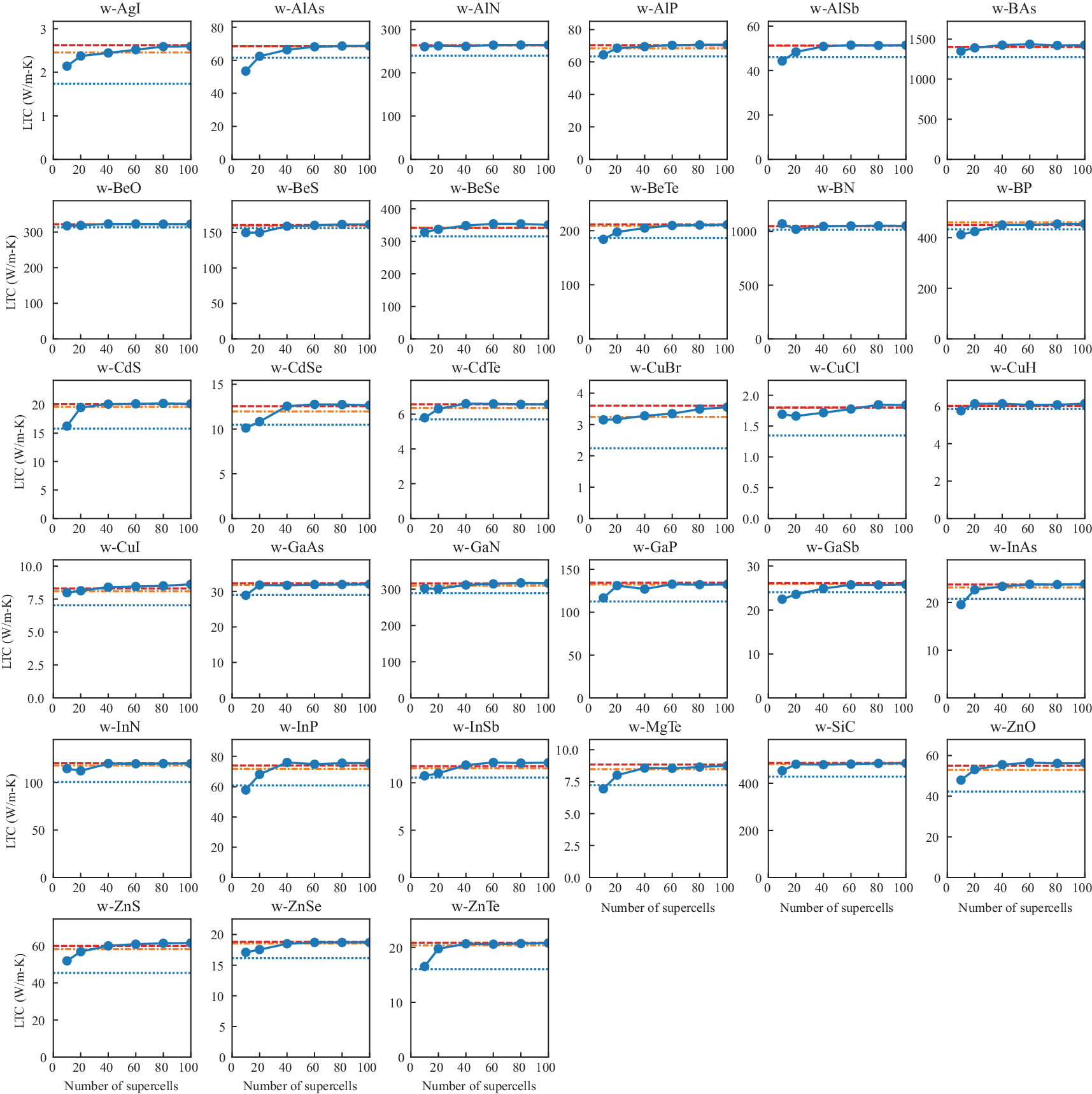}
    \caption{\label{fig:wrutzite-LTC} Filled circles show LTCs ($\kappa$) of the
    33 wurtzite-type compounds calculated at 300 K with respect to the number of
    supercells in the datasets used to train the polynomial MLPs. The LTC values
    are the averages of the diagonal elements, i.e., $(2\kappa_{xx} +
    \kappa_{zz})/3$. The horizontal dotted, dashed-dotted, and dashed lines
    depict the LTC values calculated in the conventional approach from the
    datasets of 100, 400, and 2000 supercells
    without using the polynomial MLPs, respectively.}
  \end{center}
\end{figure*}

\begin{figure*}[ht]
  \begin{center}
    \includegraphics[width=1.0\linewidth]{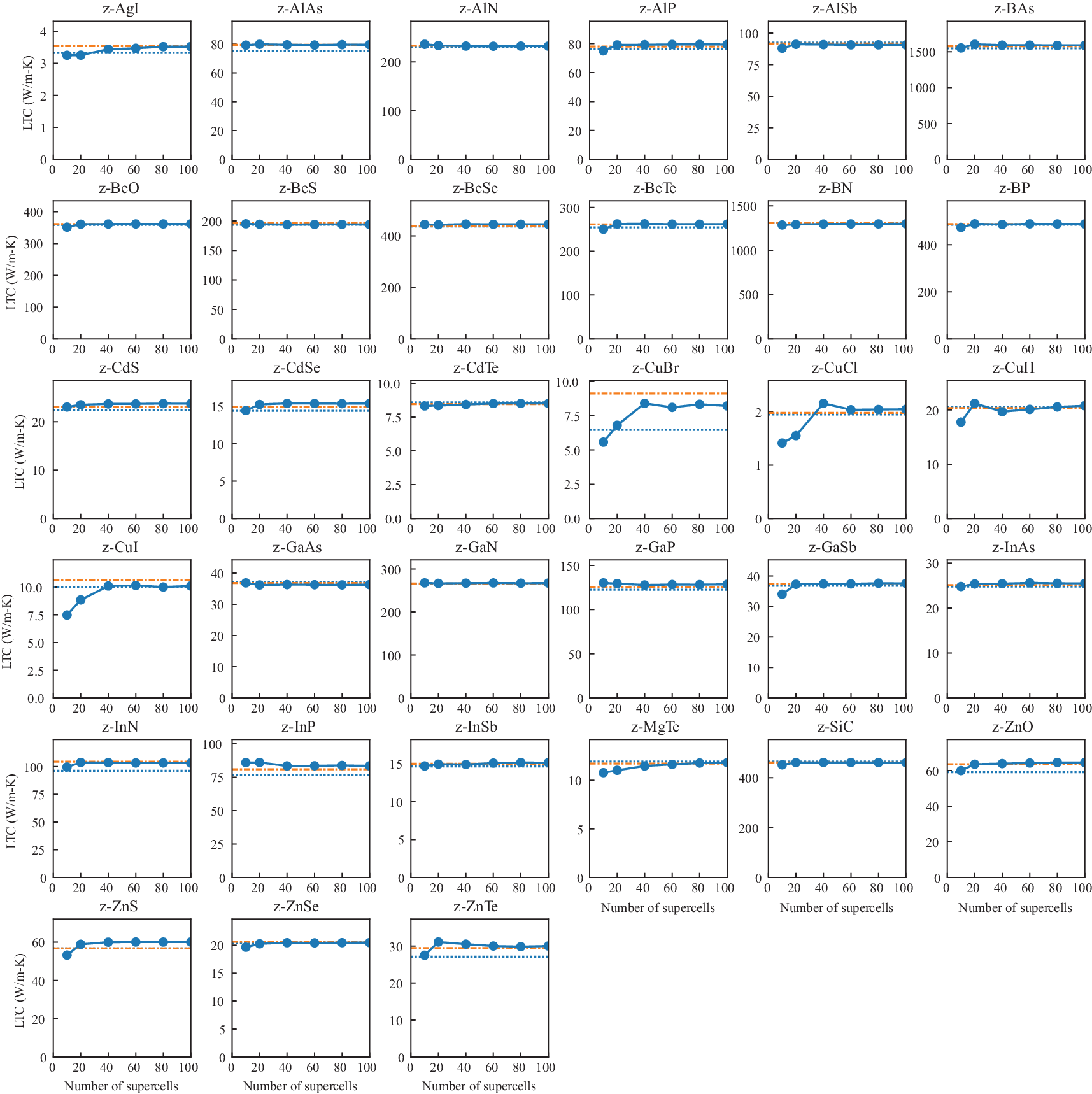}
    \caption{\label{fig:zincblende-LTC} Filled circles show LTCs of the 33
    zincblende-type compounds calculated at 300 K with respect to the number of
    supercells in the datasets used to train the polynomial MLPs. The horizontal
    dotted and dashed-dotted lines depict the LTC values calculated in the
    conventional approach from the datasets of 100 and 400 supercells
    without using the polynomial MLPs, respectively.}
  \end{center}
\end{figure*}

\begin{figure*}[ht]
  \begin{center}
    \includegraphics[width=1.0\linewidth]{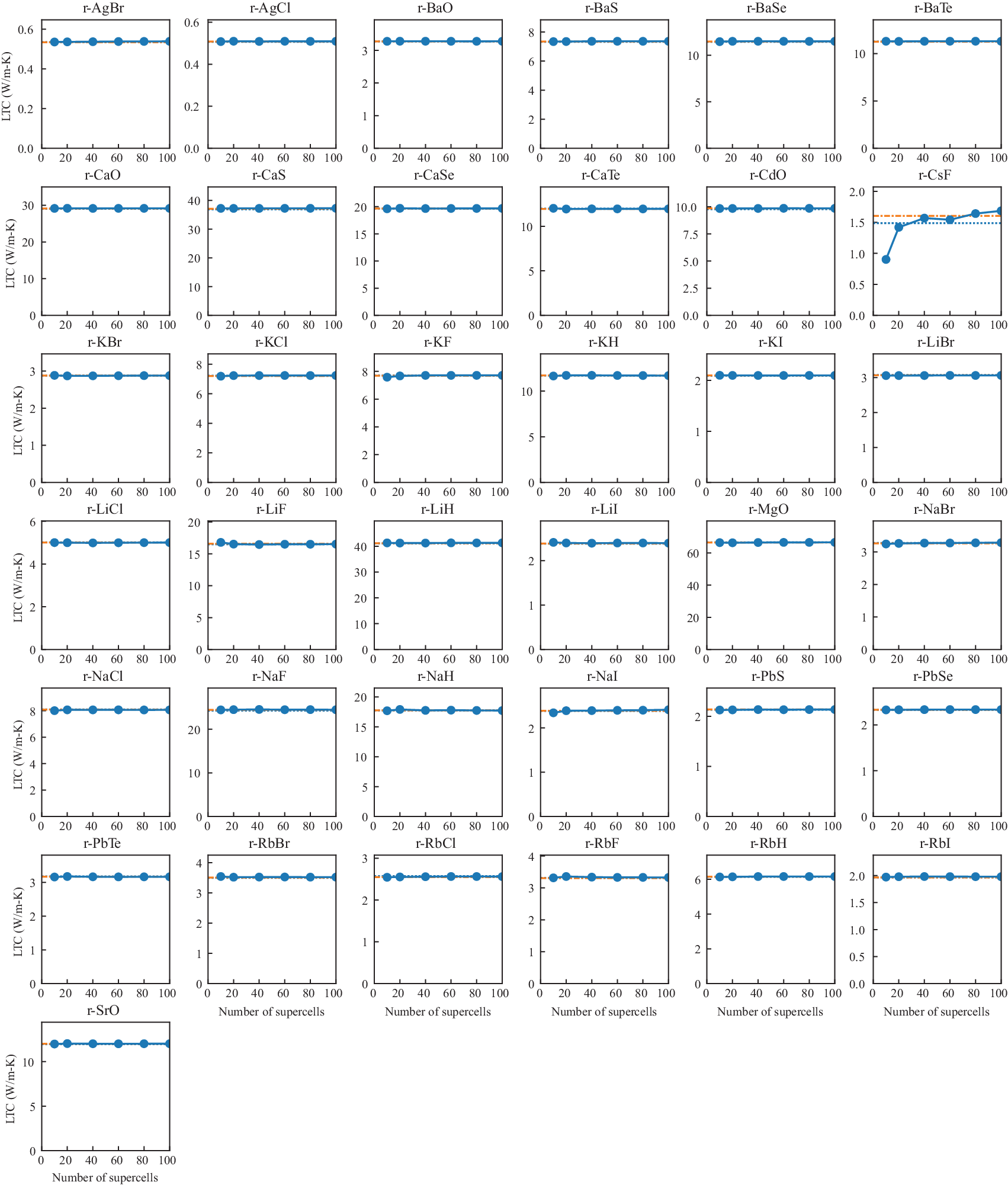}
    \caption{\label{fig:rocksalt-LTC}
    Filled circles show LTCs of the 37 rocksalt-type compounds calculated at 300
    K with respect to the number of supercells in the datasets used to train the
    polynomial MLPs. The horizontal dotted and dashed-dotted lines depict the
    LTC values calculated in the conventional approach from the datasets of 100
    and 400 supercells without using the polynomial MLPs, respectively.}
  \end{center}
\end{figure*}

In Figs.~\ref{fig:wrutzite-LTC}, \ref{fig:zincblende-LTC}, and
\ref{fig:rocksalt-LTC}, we present the calculated LTCs of the 103 compounds at
300 K. We can see that datasets with 20 supercells show good performance, at
least for estimating LTC values roughly. In particular, the LTC values of most
of the rocksalt-type compounds are well represented by these small datasets. The
wurtzite- and zincblende-type compounds exhibit similar tendencies in LTC values
with respect to dataset size since these crystal structures are similar. The
datasets with 40 supercells yield LTC results that are roughly converged.

LTC values at 300 K predicted by the conventional approach, which directly uses
the displacement-force dataset obtained through first-principles calculations to
fit third-order supercell force constants, are depicted by the horizontal dotted
lines. The third-order supercell force constants were computed by the linear
regression method as implemented in the symfc code~\cite{symfc-project} from the
first datasets with 100 supercells and 0.03 \AA~random directional
displacements, which were those prepared for training the polynomial MLPs, as
explained in Sec.~\ref{sec:choice-of-disps}. In addition, LTC values with 400
supercells for the zincblende- and rocksalt-type compounds and those with 400
and 2000 supercells for the wurtzite-type compounds were also computed.
These values are depicted as horizontal lines in Figs.~\ref{fig:wrutzite-LTC},
\ref{fig:zincblende-LTC}, and \ref{fig:rocksalt-LTC}. For most of the
zincblende- and rocksalt-type compounds, LTC values derived from datasets with
100 supercells are found to be adequate when compared to those from 400
supercells. However, for the wurtzite-type compounds, even datasets with 400
supercells are insufficient.

The LTC values predicted for the wurtzite-type compounds using polynomial
MLPs tend to align with those calculated directly
from 2000 supercell datasets. This alignment emphasizes the utility and
effectiveness of using polynomial MLPs in these cases.

\subsection{Comparison with conventional LTC calculation}

\begin{figure}[ht]
  \begin{center}
    \includegraphics[width=1.00\linewidth]{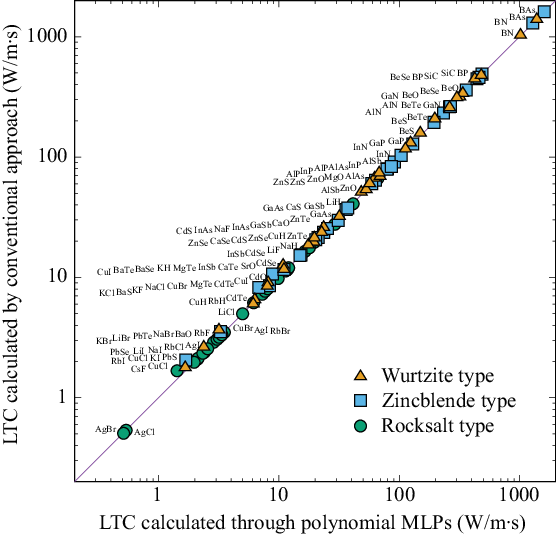}
    \caption{\label{fig:ltc-comp}
    Comparison between LTC values calculated through the polynomial MLPs and
    those by the conventional approach.\cite{phonopy-phono3py-JPSJ} These values
    are tabulated in Tables~\ref{table:zincblende-wurtzite-kappa} and
    \ref{table:rocksalt-kappa}. To train the polynomial MLPs for each compound,
    we employed displacement-force pairs and energies of 20 supercells obtained
    through first-principles calculations. For the latter LTCs, we used the
    finite-difference displacement-force
    datasets~\cite{phono3py-fd-103compounds} from
    Ref.~\onlinecite{phonopy-phono3py-JPSJ}, comprising 1254, 222, and 146
    supercells for the wurtzite-, zincblende-, and rocksalt-type compounds,
    respectively, to fit third-order supercell force constants.}
  \end{center}
\end{figure}

In Fig.~\ref{fig:ltc-comp}, the LTC values of the 103 compounds calculated
through the polynomial MLPs trained using the 20 supercell datasets are compared
with those calculated in the conventional approach using the same
finite-difference displacement-force datasets~\cite{phono3py-fd-103compounds} as
those employed in Ref.~\onlinecite{phonopy-phono3py-JPSJ}. These datasets share
the same unit cells and supercell sizes for each compound. The latter datasets
for the wurtzite-, zincblende-, and rocksalt-type compounds consist of 1254,
222, and 146 displacements, respectively, with a displacement distance of 0.03
\AA. These displacements were systematically introduced considering crystal
symmetries~\cite{Laurent-phph-2011} by using the phono3py code.\cite{phono3py,
phonopy-phono3py-JPCM} In all these calculations, the same version of the
phono3py code~\cite{phono3py, phonopy-phono3py-JPCM} (release v3.0.3) was
utilized to calculate the LTCs from the respective supercell force constants.
The results demonstrate that the LTC values obtained through the polynomial MLPs
consistently agree with those predicted by the conventional
approach.\cite{{phonopy-phono3py-JPSJ}} The LTC values of the 103 compounds are
also tabulated in Tables~\ref{table:zincblende-wurtzite-kappa} and
\ref{table:rocksalt-kappa}.

\begin{table}[ht]
  \caption{\label{table:zincblende-wurtzite-kappa} LTC values of zincblende- and
  wurtzite-type compounds at 300 K in W/m$\cdot$s as shown in
  Fig.~\ref{fig:ltc-comp} calculated through the
  polynomial MLPs, z-$\kappa^\text{pMLP}$ and w-$\kappa^\text{pMLP}$,
  respectively, and those by the conventional approach, z-$\kappa^\text{conv}$
  and z-$\kappa^\text{conv}$, respectively.
  }
  \begin{ruledtabular}
    \begin{tabular}{lrrrrrr}
      & z-$\kappa^\text{conv}$ & z-$\kappa^\text{pMLP}$ & w-$\kappa^\text{conv}_{xx}$
      & w-$\kappa^\text{conv}_{zz}$ & w-$\kappa^\text{pMLP}_{xx}$ & w-$\kappa^\text{pMLP}_{zz}$  \\
      \hline
      AgI & 3.538 & 3.260 & 2.491 & 3.008 & 2.229 & 2.654 \\
      AlAs & 79.84 & 79.88 & 68.18 & 66.24 & 62.99 & 61.21 \\
      AlN & 232.0 & 233.6 & 269.5 & 243.7 & 271.4 & 244.4 \\
      AlP & 79.18 & 79.21 & 71.84 & 65.52 & 71.17 & 65.54 \\
      AlSb & 91.03 & 91.19 & 52.59 & 49.54 & 49.09 & 46.89 \\
      BAs & 1614 & 1605 & 1535 & 1155 & 1541 & 1092 \\
      BeO & 361.7 & 361.0 & 317.8 & 325.0 & 316.8 & 322.9 \\
      BeS & 194.2 & 194.4 & 165.7 & 149.1 & 153.6 & 141.8 \\
      BeSe & 447.3 & 443.3 & 345.7 & 333.7 & 342.2 & 327.8 \\
      BeTe & 262.3 & 262.3 & 214.4 & 202.3 & 200.5 & 191.6 \\
      BN & 1301 & 1290 & 1058 & 1012 & 1043 & 972.2 \\
      BP & 489.5 & 488.3 & 498.5 & 359.9 & 468.0 & 338.1 \\
      CdS & 23.67 & 23.52 & 19.81 & 19.99 & 19.40 & 19.71 \\
      CdSe & 15.38 & 15.26 & 12.53 & 13.19 & 10.61 & 11.31 \\
      CdTe & 8.523 & 8.358 & 6.284 & 7.215 & 6.043 & 6.812 \\
      CuBr & 8.219 & 6.795 & 3.177 & 4.711 & 2.832 & 3.845 \\
      CuCl & 2.056 & 1.681 & 1.452 & 2.483 & 1.383 & 2.224 \\
      CuH & 21.44 & 21.16 & 6.248 & 5.724 & 6.202 & 5.949 \\
      CuI & 10.68 & 8.844 & 8.773 & 9.872 & 7.777 & 8.821 \\
      GaAs & 36.66 & 36.19 & 33.52 & 30.27 & 32.68 & 30.01 \\
      GaN & 263.1 & 266.7 & 318.0 & 308.3 & 306.9 & 289.8 \\
      GaP & 128.4 & 129.4 & 139.7 & 118.9 & 130.9 & 110.7 \\
      GaSb & 37.74 & 37.30 & 28.5 & 21.90 & 25.25 & 20.13 \\
      InAs & 25.47 & 25.35 & 24.23 & 23.28 & 22.92 & 22.08 \\
      InN & 103.3 & 103.7 & 116.8 & 120.6 & 112.1 & 113.2 \\
      InP & 83.65 & 85.97 & 75.73 & 72.38 & 70.61 & 63.27 \\
      InSb & 15.21 & 14.96 & 11.76 & 11.84 & 10.96 & 11.08 \\
      MgTe & 11.65 & 11.04 & 8.263 & 9.050 & 7.834 & 8.341 \\
      SiC & 459.9 & 460.8 & 512.6 & 426.4 & 508.4 & 425.1 \\
      ZnO & 64.27 & 63.47 & 51.38 & 59.91 & 50.80 & 57.65 \\
      ZnS & 60.17 & 58.97 & 59.81 & 63.11 & 55.63 & 59.30 \\
      ZnSe & 20.55 & 20.25 & 18.77 & 18.88 & 17.61 & 17.39 \\
      ZnTe & 29.79 & 31.08 & 21.54 & 21.43 & 19.78 & 19.56 \\
    \end{tabular}
  \end{ruledtabular}
\end{table}

\begin{table}[ht]
  \caption{\label{table:rocksalt-kappa} LTC values of rocksalt-type compounds at
  300 K in W/m$\cdot$s shown in Fig.~\ref{fig:ltc-comp} calculated through the
  polynomial MLPs (r-$\kappa^\text{pMLP}$) and those by the conventional
  approach (r-$\kappa^\text{conv}$). }
  \begin{ruledtabular}
    \begin{tabular}{lrr}
      & r-$\kappa^\text{conv}$ & r-$\kappa^\text{pMLP}$ \\
      \hline
      AgBr & 0.5341 & 0.5363 \\
      AgCl & 0.5053 & 0.5095 \\
      BaO & 3.267 & 3.280 \\
      BaS & 7.348 & 7.347 \\
      BaSe & 11.48 & 11.50 \\
      BaTe & 11.28 & 11.30 \\
      CaO & 27.52 & 29.18 \\
      CaS & 37.10 & 37.23 \\
      CaSe & 19.56 & 19.70 \\
      CaTe & 11.92 & 11.94 \\
      CdO & 9.794 & 9.845 \\
      CsF & 1.668 & 1.420 \\
      KBr & 2.880 & 2.869 \\
      KCl & 7.208 & 7.230 \\
      KF & 7.648 & 7.670 \\
      KH & 11.58 & 11.72 \\
      KI & 2.087 & 2.098 \\
      LiBr & 3.057 & 3.058 \\
      LiCl & 4.972 & 4.995 \\
      LiF & 16.50 & 16.53 \\
      LiH & 40.86 & 41.25 \\
      LiI & 2.375 & 2.397 \\
      MgO & 66.35 & 66.35 \\
      NaBr & 3.268 & 3.263 \\
      NaCl & 8.030 & 8.070 \\
      NaF & 24.39 & 24.47 \\
      NaH & 17.62 & 17.90 \\
      NaI & 2.409 & 2.389 \\
      PbS & 2.130 & 2.129 \\
      PbSe & 2.331 & 2.331 \\
      PbTe & 3.163 & 3.175 \\
      RbBr & 3.501 & 3.518 \\
      RbCl & 2.542 & 2.553 \\
      RbF & 3.330 & 3.357 \\
      RbH & 6.145 & 6.141 \\
      RbI & 1.967 & 1.978 \\
      SrO & 11.98 & 12.03 \\
    \end{tabular}
  \end{ruledtabular}
\end{table}

\subsection{Conclusion}
To improve the efficiency of high-throughput LTC calculations, we developed
methodologies and modular software packages that utilize polynomial MLPs for
computing LTC values based on first-principles calculation. We evaluated the
feasibility of this computational approach by calculating the LTCs of 103
compounds of wurtzite, zincblende, and rocksalt types. This approach was
benchmarked against our previously used conventional approach. We found that
this approach significantly reduces computational demands while maintaining a
satisfactory accuracy level for LTC prediction. Apart from the initial stage of
generating datasets using first-principles calculations, subsequent LTC
calculation steps require minimal computational resources. This enables users to
calculate LTCs and various related physical values on standard computers, given
access to high-quality datasets. Our future plans include the computation and
distribution of such high-quality datasets for a wide range of compounds.

\begin{acknowledgments}
This work was supported by JSPS KAKENHI Grant Numbers JP21K04632, JP22H01756,
JP19H05787 and 24K08021 Some of the calculations in this study were performed on
the Numerical Materials Simulator at NIMS.
\end{acknowledgments}

\bibliography{JCP}
\end{document}